\documentclass[runningheads]{llncs}

\usepackage{iris}
\usepackage{pftools}

\usepackage[T1]{fontenc}
\usepackage[utf8]{inputenc}

\usepackage{inconsolata}

\usepackage{amsmath}
\usepackage{amssymb}
\usepackage{amsfonts}

\usepackage{listings, listings-rust}

\usepackage{simplebnf}

\SetBNFConfig{
    relation = {->|:in:|:ni:|::=|:emp:|:eqdef:|:eq:},
    relation-sym-map =
        {
            {->} = {\ensuremath{\rightarrow}},
            {:in:} = {\ensuremath{\in}},
            {:ni:} = {\ensuremath{\ni}},
            {::=} = {\ensuremath{\Coloneqq}},
            {:emp:} = {},
            {:eqdef:} = {\ensuremath{\eqdef}},
            {:eq:} = {\ensuremath{=}},
    },
}

\usepackage{mathpartir}

\usepackage{marvosym}
\pretocmd\mvchr{\text}{}{\errmessage{Patching \noexpand\mvchr failed}}
\pretocmd\textmvs{\text}{}{\errmessage{Patching \noexpand\textmvs failed}}

\usepackage{newclude}

\usepackage{bm}

\usepackage[framemethod=TikZ]{mdframed}

\usepackage{verbatim}

\usepackage{color}
\usepackage{hyperref}
\hypersetup{
    colorlinks=true,
    urlcolor=blue,
    linkcolor=cyan,
    plainpages=false,
    pdfusetitle=true,
}
\usepackage{fontawesome5}

\usepackage{wrapfig}

\newcommand{\coreCalculus}{\ensuremath{\lambda_{\mathit{PA}}}}

\newcommand{\Hoaretriple}[3]{%
  \{\,#1\,\}
  \mathrel{#2}\nolinebreak
  \{\,#3\,\}
}

\newcommand{\sepconj}{\ensuremath{\bm{\ast{}}}}

\DeclareMathOperator*{\qmapsto}{\mapsto}
\newcommand{\pointsto}[2]{#1 \ensuremath{\mapsto} #2} 
\newcommand{\pointstofrac}[3]{\ensuremath{#1 \qmapsto^{#3} #2}} 
\newcommand{\pointstofracinline}[3]{#1 \ensuremath{\qmapsto\limits^{#3}} #2}
\newcommand{\pointstolim}[2]{\ensuremath{\qmapsto\limits^{\frac{#1}{#2}}}}

\DeclareMathOperator*{\Refend}{\Uparrow}
\newcommand{\refend}[3]{#1 \ensuremath{\Refend^{#3}} #2}
\newcommand{\refendinline}[3]{#1 \ensuremath{\Refend\limits^{#3}} #2}
\newcommand{\refendop}[2]{\ensuremath{\Refend\limits^{\frac{#1}{#2}}}}

\DeclareMathOperator*{\Resend}{\Downarrow}

\newcommand{\emp}{\textbf{emp}}

\newcommand{\pure}[1]{#1}
\newcommand{\deref}[1]{\texttt{*}#1}
\newcommand{\shrborrow}{\texttt{\&*}}
\newcommand{\mutborrow}{\texttt{\&mut *}}
\newcommand{\lamfun}[2]{\ensuremath{#1.\ #2}}
\newcommand{\pureExpr}{\ensuremath{\varepsilon}}

\definecolor{seporange}{rgb}{0.9, 0.5, 0.2}
\lstdefinelanguage{lambda-pa}{
    keywords = [1]{let, rec, if, then, else, fork, mut},
    keywords = [2]{new, free, retag, retag_and_protect, unprotect},
    morecomment = [l]{//},
    morecomment = [s][\color{seporange}]{ \{ }{ \} },
    mathescape,
    escapebegin=\color{seporange},
    escapeinside={(*@}{@*)},
    literate=
        {:=}{{$\coloneqq{}$}}1
        {\&*\& }{{$\sepconj{}$} }2 
        {->}{$\rightarrow$}{1}
        {|->}{{$\mapsto{}$}}1
        {resend}{$\Resend$}{1}
        {refend }{$\Refend$ }{2}
        {DANGER }{$\Biohazard$ }{2}
        {emp}{{\emp}}{3}
        {vs}{{\color{magenta}$\vs$}}{1}
}

\definecolor{codegreen}{rgb}{0,0.6,0}
\definecolor{codegray}{rgb}{0.5,0.5,0.5}
\definecolor{codepurple}{rgb}{0.58,0,0.82}
\definecolor{backcolour}{rgb}{0.95,0.95,0.92}

\lstdefinestyle{mystyle}{
    commentstyle=\color{codegreen},
    keywordstyle={[1]\bfseries},
    keywordstyle ={[2]\color{blue}\bfseries},
    numberstyle=\tiny\color{codegray},
    stringstyle=\color{codepurple},
    basicstyle=\ttfamily\footnotesize,
    breakatwhitespace=false,
    breaklines=true,
    captionpos=b,
    numbers=left,
    numbersep=5pt,
    showspaces=false,
    showstringspaces=false,
    showtabs=false,
    tabsize=2,
    xleftmargin=.14in,
    showlines=false,
}

\AtBeginEnvironment{mathparpagebreakable}{\fontsize{8}{9.5}\selectfont}

\surroundwithmdframed[
    backgroundcolor=backcolour!30,
    innerleftmargin=5pt,
    innertopmargin=2pt,
    innerrightmargin=5pt,
    innerbottommargin=2pt
]{lstlisting}
\begin{document}
    \title{Towards verifying unsafe Rust programs against Rust's pointer-aliasing restrictions}
    \titlerunning{Towards verifying Rust programs against pointer-aliasing restrictions}
    \author{Wannes Tas\inst{1}\orcidID{0009-0008-5519-761X} \and Bart Jacobs\inst{1}\orcidID{0000-0002-3605-249X}}
    \institute{Department of Computer Science, Leuven, Belgium \\ \email{firstname.lastname@kuleuven.be}}
    \maketitle

    \begin{abstract}
        The Rust programming language is famous for its strong ownership regime: at each point, each value is either exclusively owned, exclusively borrowed through a mutable reference, or borrowed as read-only through one or more shared references.
        These rules, known as Rust's pointer-aliasing rules, are exploited by the Rust compiler to generate more efficient machine code, and enforced by Rust's static type system, except inside \texttt{unsafe} blocks.
        In this paper, we present our work in progress towards the first program logic for modularly verifying that Rust programs that use unsafe blocks comply with the pointer-aliasing rules.

        \keywords{Rust \and Formal verification \and Pointer-aliasing} 
    \end{abstract}
    \section{Introduction}\label{sec:introduction}
    The Rust programming language is famous for its strong ownership regime: at each point, each value is either exclusively owned, exclusively borrowed through a mutable reference, or borrowed as read-only through one or more shared references.
    These rules, known as Rust's pointer-aliasing rules, are exploited by the Rust compiler to generate more efficient machine code, and enforced by Rust's static type system, except inside \texttt{unsafe} blocks.
    An example of a program that does not comply with Rust's aliasing restrictions can be found in Listing~\ref{lst:rust-aliasing} (L).
    The compiler should reject the second attempt to write through \texttt{y}, as the lifetime of \texttt{y} would have to be ended to write through \texttt{x} in the line before it.
    In order to verify that code in \texttt{unsafe} blocks is also compliant with Rust's pointer-aliasing rules, external tools and techniques are needed.
    To this end, we present our work in progress towards the first program logic capable of modularly verifying that Rust programs using unsafe blocks are compliant with the pointer-aliasing rules.

    As verifying arbitrary programs written in Rust's full surface syntax would be quite cumbersome, a smaller toy language is introduced in \S \ref{sec:lambda-pa}, which should be sufficient to convey the core ideas.
    Afterwards in \S \ref{sec:the-program-logic}, we present our program logic.
    Finally, a number of planned extensions of the program logic are discussed in \S \ref{sec:conclusion}, along with the most important related work.
    \begin{table*}
        \noindent
        \begin{minipage}{.45\columnwidth}
            \begin{lstlisting}[language=Rust, label={lst:rust-aliasing}]
let mut x = 42;
let y = &mut x as *mut i32;
unsafe { *y = 43; }
x = 44;
unsafe { *y = 5; } // UB!
            \end{lstlisting}
        \end{minipage}\hfill
        \begin{minipage}{.45\columnwidth}
            \begin{lstlisting}[language=lambda-pa, label={lst:lambda-pa-aliasing}]
let x = new(42) in
let y = &mut *x in
*y := 43;
*x := 44;
*y := 5
            \end{lstlisting}
        \end{minipage}
        \vspace{5pt}
        \caption{An example of a Rust program that does not satisfy Rust's pointer-aliasing restrictions (L), and its translation to $\coreCalculus$ (R).}
    \end{table*}

    \section{The mini-language \texorpdfstring{\coreCalculus{}}{lambda-PA}}\label{sec:lambda-pa}
    \begin{figure}
    \begin{bnf}[
        colspec = {rlcll},
        column{1} = {mode = dmath},
        column{2} = {mode = dmath},
        column{4} = {font = \ttfamily},
    ]
        \mathit{Provenances} : \mathit{pr} \in :eqdef: \( \mathit{OwningProv} \cup \mathit{MutRefProv} \cup \mathit{SharedRefProv} \)
        ;;
        \mathit{Pointers} : p \in  ::= \( \left\{ \texttt{prov} \colon \mathit{Provenance},\, \texttt{address} \colon \mathbb{N} \right\} \)
        ;;
        \mathit{OwningPointers} : p_\texttt{own} \in :eqdef: \( \left\{p \mid p.\texttt{prov} \in \mathit{OwningProv} \right\} \)
        ;;
        \mathit{MutRefs} : p_{\texttt{\&mut}} \in :eqdef: \( \left\{p \mid p.\texttt{prov} \in \mathit{MutRefProv} \right\} \)
        ;;
        \mathit{SharedRefs} : p_{\texttt{\&}} \in :eqdef: \( \left\{ p \mid p.\texttt{prov} \in \mathit{SharedRefProv} \right\} \)
        ;;
        \mathit{Values} : v \in ::= \( p \) // \( z \) : \( z \in \mathbb{Z} \)
        ;;
        \mathit{Variables} : x \in :emp:
        ;;
        \mathit{PureExprs} : \pureExpr \in ::= \( v \) // \( x \)
        ;;
        \mathit{Exprs} : e \in ::= \( v \)
        // new(\( \varepsilon \))
        // free(\( \pureExpr \))
        // \shrborrow{}\( \pureExpr \)
        // \mutborrow{}\( \varepsilon \)
        // \(\deref{\pureExpr}\)
        | *\( \pureExpr \coloneqq v \)
        // let \( x \) \(\coloneqq\) \( e_1\) in \( e_2 \)
        // \( e_1 \); \( e_2 \)
        ;;
    \end{bnf}
    \caption{The \coreCalculus{} syntax, inspired by \( \lambda_\mathit{Rust} \)~\cite{jung_understanding_2020} and \( \lambda^\mathit{TB} \)~\cite{peterson_program_2025}.}
    \label{fig:calculus-syntax}
\end{figure}
We define a minimal language that captures the essence of pointer-aliasing in Rust: $\coreCalculus$.
The full syntax of $\coreCalculus$ can be found in Figure~\ref{fig:calculus-syntax}.
While only the address is retained at run time, compilers take into account a pointer's provenance~\cite{the_rust_teams_stdptr_nodate} for optimisations; hence our logic must account for it.
For instance a pointer \texttt{x} points to the same memory as a borrow like \texttt{\&x} or \texttt{\&mut x}, but each would have a different provenance, since it consists, among other information, of the mutability permisions of the pointer.
Note that \texttt{$e_1$; $e_2$} is just syntactic sugar for \texttt{let \_ $\coloneqq$ $e_1$ in $e_2$}.
Separate operational semantics for $\coreCalculus$ are not provided, but the proof rules of the program logic introduced in the next section serve as axiomatic semantics.

    \section{The program logic}\label{sec:the-program-logic}
    We define the assertions of our logic semantically as predicates over fractional heaps (fractional multisets of resources).
When it is clear from context whether the assertion or the resource is meant, \( \pointstofrac{p}{v}{} \) will be used as shorthand for \( \pointstofracinline{p}{v}{1} \), and \( \refend{p'}{p}{} \) as shorthand for \( \refendinline{p'}{p}{1}\).
We define the correctness judgment inductively by the proof rules shown.

\begin{bnf}[
    colspec = {llcll},
    column{1} = {font = \sffamily},
    column{2} = {mode = dmath},
    column{4} = {mode = dmath},
]
    r : Resources ::= \pointstofrac{p}{v}{ }\
    //\ \refend{p_1}{p_2}{ }\
    ;;
    h : Heaps :in: \mathit{Resources} \rightarrow [0, 1]
    ;;
    a : Assertions :in: \mathcal{P}(\mathit{Heaps})
    ;;
    a_1 \sepconj a_2 : :eqdef: \{ h \mid \exists h_1 \in a_1, h_2 \in a_2.\; h_1 + h_2 = h\}
    ;;
    \emp : :eqdef: \left\{ \lambda r.\ 0 \right\}
    ;;
    \mathsf{own}_q\ r :eqdef: \{ (\lambda r'.\;0)[r \coloneqq q] \} : \( q \in (0, 1]\)
    ;;
    \pointstofrac{p}{v}{q} :eqdef: \mathsf{own}_q\ \pointsto{p}{v}
    ;;
    \refend{p'}{p}{q} :eqdef: \mathsf{own}_q\ \refend{p'}{p}{}
    ;;
\end{bnf}

\begin{mathparpagebreakable}
    \inferH{H-frame}{
        \Hoaretriple{P}{e}{Q}
    }{
        \Hoaretriple{R \sepconj P}{e}{Q \sepconj R}
    }
\and
    \inferH{H-exists}{
        \forall x.\ \Hoaretriple{P(x)}{e}{Q}
    }{
        \Hoaretriple{\exists x.\ P(x)}{e}{Q}
    }
\end{mathparpagebreakable}
\begin{mathparpagebreakable}
    \inferH{new}{}{
    \Hoaretriple{\emp}{\texttt{new}(v)}{\lamfun{r}{\pointsto{r}{v} \sepconj r \in \mathit{OwningPointers}}}
    }
\and
    \inferH{free}{p \in \mathit{OwningPointers}}{
        \Hoaretriple{\pointsto{p}{v}}{\texttt{free}(p)}{\emp}
    }
\and
    \inferH{write}{}{
        \Hoaretriple{\pointsto{p}{v_o}}{\texttt{*}p \coloneqq v}{\pointsto{p}{v}}
    }
\and
    \inferH{read}{}{
        \Hoaretriple{\pointstofrac{p}{v}{q}}{\deref{p}}{\lamfun{r}{\pointstofrac{p}{v}{q} \sepconj \pure{r = v}}}
    }
\and
    \inferH{let}{
        \Hoaretriple{P}{e_1}{\lamfun{r}{R(r)}} \\
        \forall v. \Hoaretriple{R(v)}{e_2[v/x]}{Q}
    }{
        \Hoaretriple{P}{\texttt{let $x$ $\coloneqq$ $e_1$ in $e_2$}}{Q}
    }
\end{mathparpagebreakable}
Above definitions and proof rules introduce a fairly straightforward separation logic~\cite{ohearn_local_2001, reynolds_separation_2002} with a fractional heap~\cite{boyland_fractional_2013, bornat_permission_2005}.
The more interesting proof rules, namely those for mutable and shared borrows, are given separately in the following two subsections.
These proof rules are each introduced alongside an example program with proof outline.

    \subsection{Mutable references}\label{subsec:mutable-references}
\hspace{0pt}
\begin{wrapfigure}[9]{r}{0.31\textwidth}
    \vspace{-10pt}
    \begin{lstlisting}[language=lambda-pa]
let x := new(42) in
{ x |-> 42 }
let y := &mut *x in
{ y |-> 42 &*& y refend x }
*y := 43;
{ y |-> 43 &*& y refend x }
vs { x |-> 43 }
*x := 44;
{ x |-> 44  }
free(x)
    \end{lstlisting}
\end{wrapfigure}
\begin{mathparpagebreakable}
    \inferH{mutable-borrow}{}{
        \Hoaretriple{\pointsto{p}{v}
        }{
            \mutborrow{}p
        }{
            \lamfun{r}{\pointstofrac{r}{v}{} \sepconj \refend{r}{p}{}}
        }
    }
\end{mathparpagebreakable}

Creating a mutable reference in our language requires the full points-to resource of the lending reference, and produces a full points-to resource to the same memory for the borrowing pointer.
Additionally, a reference-ending resource is produced to allow the proof author to transfer (full or partial) ownership of the memory from the borrowing reference back to the lending reference, using a view shift~\cite{jung_iris_2015}:

\begin{mathparpagebreakable}
    \inferH{H-view-shift-consequence}{
        P \vs P' \\
        \Hoaretriple{P'}{e}{Q} \\
        Q \vs Q'
    }{
        \Hoaretriple{P}{e}{Q'}
    }
\and
    \inferH{VS-transitive}{
        P \vs R \\
        R \vs Q
    }{
        P \vs Q
    }
    \and
    \inferH{VS-frame}{
        P \vs Q
    }{
        R \sepconj P \vs Q \sepconj R
    }
    \and
    \inferH{VS-subset}{
        P \subseteq Q
    }{
        P \vs Q
    }
%
\and
    \inferH{reference-end}{}{
        \pointstofrac{p'}{v}{q} \sepconj \refend{p'}{p}{q} \vs \pointstofrac{p}{v}{q}
    }
\end{mathparpagebreakable}
Of course, the goal is for this to be sound with respect to Rust's aliasing rules.
The Rust Reference states the following about mutable references\footnote{\href{https://doc.rust-lang.org/reference/behavior-considered-undefined.html\#r-undefined.alias}{https://doc.rust-lang.org/reference/behavior-considered-undefined.html\#r-undefined.alias}}: ``\texttt{\&mut T} must point to memory that is not read or written by any pointer not derived from the reference and that no other reference points to while they are live.''\footnote{Interpreted literally, a choice of $q \neq 1$ in \ruleref{reference-end} would not be allowed by the Rust Reference. On the other hand, $p'$ can never be mutated again, and Tree Borrows~\cite{villani_tree_2025} allows it, so it is unclear whether this is actually an issue or not.}

\subsection{Shared References}\label{subsec:shared-references}
\begin{mathparpagebreakable}
    \inferH{shared-borrow}{}{
        \Hoaretriple{\pointstofrac{p}{v}{q}}{\shrborrow{}p}{
            \lamfun{r}{\pointstofrac{p}{v}{\frac{q}{2}} \sepconj \pointstofrac{r}{v}{\frac{q}{2}} \sepconj \refend{r}{p}{\frac{q}{2}}}
        }
    }
\end{mathparpagebreakable}
Creating a shared reference is somewhat similar to creating a mutable one, except that the lending pointer only needs to have partial ownership of its memory, and both the lending and borrowing references are read-only afterwards.
This satisfies Rust's aliasing rules for shared pointers, which are as follows\footnote{\href{https://doc.rust-lang.org/reference/behavior-considered-undefined.html\#r-undefined.alias}{https://doc.rust-lang.org/reference/behavior-considered-undefined.html\#r-undefined.alias}}: ``\texttt{\&T} must point to memory that is not mutated while they are live (except for data inside an \texttt{UnsafeCell<U>})''\footnote{\texttt{UnsafeCell<T>} is not yet considered in our program logic, see~\S \ref{sec:conclusion}.}.
Given that \ruleref{shared-borrow} only gives partial ownership to the lending pointer, and offers no way to acquire full ownership, it can never be possible to apply \ruleref{write} using this pointer, and thus mutation is not possible while the shared reference is live.
Note that in the current system only the memory directly behind a shared reference is immutable; memory that can be reached transitively through the reference is not.
See \S \ref{sec:conclusion} for more on this.
\begin{lstlisting}[language=lambda-pa,label={lst:shared-ref}]
let x := new(42) in
{ x |-> 42 }
let y := &*x in
{ x $\pointstolim{1}{2}$ 42 &*& (y $\pointstolim{1}{2}$ 42 &*&  y $\refendop{1}{2}$ x) }
let z := &*x in
{ x $\pointstolim{1}{4}$ 42 &*& (y $\pointstolim{1}{2}$ 42 &*&  y $\refendop{1}{2}$ x) &*& (z $\pointstolim{1}{4}$ 42 &*& z $\refendop{1}{4}$ x)}
*y;
vs { x $\pointstolim{3}{4}$ 42 &*& (z $\pointstolim{1}{4}$ 42 &*& z $\refendop{1}{4}$ x)}
*z;
vs { x |-> 42 }
*x := 43
{ x |-> 43 }
free(x)
\end{lstlisting}

    \section{Conclusion}\label{sec:conclusion}
    In this paper we propose a logic that captures the core of Rust's pointer aliasing rules.
In a technical report~\cite{tas_towards_2026}, we present an extended version of our logic that additionally supports two-phase borrowing~\cite{villani_tree_2025}, and the following features:

\subsubsection{Functions}
Rust's aliasing rules note an extra constraint for references that are passed as function arguments: ``When a reference (but not a \texttt{Box}!) is passed to a function, it is live at least as long as that function call, again except if the \texttt{\&T} contains an \texttt{UnsafeCell<U>}.''
Additionally, it must be verified on function entrance that mutable reference arguments are not aliasing, which is not guaranteed by Rust in the presence of \texttt{unsafe} code at the call-site.
To this end, we reborrow mutable references on function entrance, and make the reference-ending resources of reference arguments inaccessible for the duration of the function call.

\subsubsection{Interior Mutability}
While in most cases the memory behind a shared reference should be immutable, sometimes it is desirable to allow mutation through shared references, e.g.\ so multiple threads can mutate some shared memory.
For this reason Rust has the type \texttt{UnsafeCell<T>}, which is in essence just the type \texttt{T} except for the fact that it can be mutated through shared pointers.
In order to integrate this feature into the proposed logic, care must be taken such that taking a mutable or shared borrow of a type that contains or points to an \texttt{UnsafeCell<T>}, always leaves the full points-to resource intact.

\subsubsection{Boxes}
According to Rust's aliasing rules, \texttt{Box<T>} should be treated similarly to a $\texttt{\&'static mut T}$ for aliasing purposes, except for the fact that a \texttt{Box} passed as a function argument is not required to remain live for the duration of the function execution.

Additionally, the following extensions are left as future work:

\subsubsection{Structs}
Another important feature that is currently missing is the ability to define and use structs.
One significant aspect of structs is that it allows programmers to introduce abstractions, so that client code can use e.g. a \texttt{Coordinate} struct without having to worry about its internal workings.
Similarly, an addition of structs to our program logic should allow modular reasoning and verification of client code against a separately verified specification of that struct, rather than worrying about its internals.

\subsubsection{Recursive Freezing}\label{subsubsec:recursive-freezing}
In the current version of our logic, it is possible to write through a pointer of type \texttt{\&\&mut T} by first reading from it to get a pointer of type \texttt{\&mut T}, and then writing through that one.
However, the Rust Reference states the following:
``Moreover, the bytes pointed to by a shared reference, including transitively through other references (both shared and mutable) and Boxes, are immutable; transitivity includes those references stored in fields of compound types.''
To support this concept, the creation of a shared reference could recursively freeze all memory behind it so that all transitively reachable memory becomes read-only.
Note that Tree Borrows\footnote{See \S \ref{subsec:related-work}} does not require recursive freezing, so it seems quite likely that this requirement might be relaxed in the future.

Lastly, we plan on validating the logic extensively, by implementing it in the VeriFast~\cite{jacobs_featherweight_2015} tool for separation-logic-based modular formal verification of Rust programs and proving compliance of some real world Rust programs, and by proving soundness of the logic with respect to some formal model of Rust's aliasing rules like Tree Borrows.

\subsection{Related Work}\label{subsec:related-work}
We are not aware of any existing logic for modularly verifying compliance with Rust's pointer aliasing rules.
As far as we are aware, existing approaches for modularly verifying \texttt{unsafe} Rust code, such as RefinedRust~\cite{gaher_refinedrust_2024}, Verus~\cite{lattuada_verus_2023}, and Gillian-Rust~\cite{ayoun_hybrid_2025}, ignore these rules and are therefore unsound.
Some non-modular tools, however, like testing tool Miri~\cite{jung_miri_2026} and bounded model checker Kani~\cite{vanhattum_verifying_2022} do check these rules to some extent.

\subsubsection{Tree Borrows}
The official pointer-aliasing rules currently have two relevant issues: they are not formally defined, so at best our program logic verifies compliance with \emph{our interpretation} of the aliasing rules, and secondly they are not final.
The current aliasing rules are an overapproximation of what the final official aliasing rules will be, and are likely to be relaxed in some ways eventually.
A few candidate formal proposals for these have been made so far.
A recent proposal that has gained some traction is Tree Borrows~\cite{villani_tree_2025}, which keeps track of all references to some block in memory via a tree-structure, where each node represents a pointer to (a part of) that memory.
Each node is also equipped with a state machine that tracks what operations (read/write) the pointer should allow.
A separation logic to verify compliance with Tree Borrows is currently in development~\cite{peterson_program_2025}.

\begin{credits}
    \subsubsection{\ackname}
    This research was partially funded by the Research Fund KU Leuven, and by the Cybersecurity Research Program Flanders.

    We would also like to thank Johannes Hostert and Ralf Jung for their insight and helpful interactions.

    \subsubsection{\discintname}
    None
\end{credits}

    \bibliographystyle{splncs04}
    \bibliography{main}

@phdthesis{jung_understanding_2020,
	type = {{doctoralThesis}},
	title = {Understanding and evolving the {Rust} programming language},
	url = {https://publikationen.sulb.uni-saarland.de/handle/20.500.11880/29647},
	doi = {10.22028/D291-31946},
	language = {en},
	urldate = {2025-09-30},
	school = {Saarländische Universitäts- und Landesbibliothek},
	author = {Jung, Ralf},
	year = {2020},
	note = {Accepted: 2020-09-09T07:57:28Z},
}

@article{villani_tree_2025,
	title = {Tree {Borrows}},
	volume = {9},
	url = {https://dl.acm.org/doi/10.1145/3735592},
	doi = {10.1145/3735592},
	number = {PLDI},
	urldate = {2025-10-03},
	journal = {Tree Borrows -- Artifact},
	author = {Villani, Neven and Hostert, Johannes and Dreyer, Derek and Jung, Ralf},
	month = jun,
	year = {2025},
	pages = {188:1019--188:1042},
}

@mastersthesis{peterson_program_2025,
	title = {A {Program} {Logic} for {Tree} {Borrows}},
	url = {http://rudynicolop.github.io/files/thesis.pdf},
	school = {ETH Zurich},
	author = {Peterson, Rudy},
	month = sep,
	year = {2025},
}

@article{gaher_refinedrust_2024,
	title = {{RefinedRust}: {A} {Type} {System} for {High}-{Assurance} {Verification} of {Rust} {Programs}},
	volume = {8},
	shorttitle = {{RefinedRust}},
	url = {https://dl.acm.org/doi/10.1145/3656422},
	doi = {10.1145/3656422},
	number = {PLDI},
	urldate = {2025-10-30},
	journal = {Artifact for "RefinedRust: A Type System for High-Assurance Verification of Rust Programs"},
	author = {Gäher, Lennard and Sammler, Michael and Jung, Ralf and Krebbers, Robbert and Dreyer, Derek},
	month = jun,
	year = {2024},
	pages = {192:1115--192:1139},
}

@inproceedings{reynolds_separation_2002,
	title = {Separation logic: a logic for shared mutable data structures},
	issn = {1043-6871},
	shorttitle = {Separation logic},
	url = {https://ieeexplore.ieee.org/document/1029817},
	doi = {10.1109/LICS.2002.1029817},
	urldate = {2025-11-20},
	booktitle = {Proceedings 17th {Annual} {IEEE} {Symposium} on {Logic} in {Computer} {Science}},
	author = {Reynolds, J.C.},
	month = jul,
	year = {2002},
	pages = {55--74},
}

@incollection{boyland_fractional_2013,
	address = {Berlin, Heidelberg},
	title = {Fractional {Permissions}},
	isbn = {978-3-642-36946-9},
	url = {https://doi.org/10.1007/978-3-642-36946-9_10},
	doi = {10.1007/978-3-642-36946-9_10},
	language = {en},
	urldate = {2026-01-08},
	booktitle = {Aliasing in {Object}-{Oriented} {Programming}. {Types}, {Analysis} and {Verification}},
	publisher = {Springer},
	author = {Boyland, John},
	editor = {Clarke, Dave and Noble, James and Wrigstad, Tobias},
	year = {2013},
	pages = {270--288},
}

@article{bornat_permission_2005,
	title = {Permission accounting in separation logic},
	volume = {40},
	issn = {0362-1340},
	url = {https://dl.acm.org/doi/10.1145/1047659.1040327},
	doi = {10.1145/1047659.1040327},
	number = {1},
	urldate = {2026-01-08},
	journal = {SIGPLAN Not.},
	author = {Bornat, Richard and Calcagno, Cristiano and O'Hearn, Peter and Parkinson, Matthew},
	month = jan,
	year = {2005},
	pages = {259--270},
}

@article{jung_iris_2015,
	title = {Iris: {Monoids} and {Invariants} as an {Orthogonal} {Basis} for {Concurrent} {Reasoning}},
	volume = {50},
	issn = {0362-1340},
	shorttitle = {Iris},
	url = {https://dl.acm.org/doi/10.1145/2775051.2676980},
	doi = {10.1145/2775051.2676980},
	number = {1},
	urldate = {2026-01-08},
	journal = {SIGPLAN Not.},
	author = {Jung, Ralf and Swasey, David and Sieczkowski, Filip and Svendsen, Kasper and Turon, Aaron and Birkedal, Lars and Dreyer, Derek},
	month = jan,
	year = {2015},
	pages = {637--650},
}

@misc{the_rust_teams_stdptr_nodate,
	title = {std::ptr - {Rust}},
	url = {https://doc.rust-lang.org/std/ptr/index.html#provenance},
	urldate = {2026-01-12},
	author = {{The Rust teams}},
}

@inproceedings{ohearn_local_2001,
	address = {Berlin, Heidelberg},
	title = {Local {Reasoning} about {Programs} that {Alter} {Data} {Structures}},
	isbn = {978-3-540-44802-0},
	doi = {10.1007/3-540-44802-0_1},
	language = {en},
	booktitle = {Computer {Science} {Logic}},
	publisher = {Springer},
	author = {O’Hearn, Peter and Reynolds, John and Yang, Hongseok},
	editor = {Fribourg, Laurent},
	year = {2001},
	pages = {1--19},
}

@article{jacobs_featherweight_2015,
	title = {Featherweight {VeriFast}},
	volume = {Volume 11, Issue 3},
	issn = {1860-5974},
	url = {https://lmcs.episciences.org/1595},
	doi = {10.2168/LMCS-11(3:19)2015},
	urldate = {2026-01-17},
	journal = {Logical Methods in Computer Science},
	publisher = {Episciences.org},
	author = {Jacobs, Bart and Vogels, Frédéric and Piessens, Frank},
	month = sep,
	year = {2015},
}

@article{lattuada_verus_2023,
	title = {Verus: {Verifying} {Rust} {Programs} using {Linear} {Ghost} {Types}},
	volume = {7},
	shorttitle = {Verus},
	url = {https://dl.acm.org/doi/10.1145/3586037},
	doi = {10.1145/3586037},
	number = {OOPSLA1},
	urldate = {2026-01-17},
	journal = {Software Artifact (virtual machine, pre-built distributions) for "Verus: Verifying Rust Programs using Linear Ghost Types"},
	author = {Lattuada, Andrea and Hance, Travis and Cho, Chanhee and Brun, Matthias and Subasinghe, Isitha and Zhou, Yi and Howell, Jon and Parno, Bryan and Hawblitzel, Chris},
	month = apr,
	year = {2023},
	pages = {85:286--85:315},
}

@article{ayoun_hybrid_2025,
	title = {A {Hybrid} {Approach} to {Semi}-automated {Rust} {Verification}},
	volume = {9},
	url = {https://dl.acm.org/doi/10.1145/3729289},
	doi = {10.1145/3729289},
	number = {PLDI},
	urldate = {2026-01-17},
	journal = {Artifact: A Hybrid Approach to Semi-automated Rust Verification},
	author = {Ayoun, Sacha-{\'E}lie and Denis, Xavier and Maksimović, Petar and Gardner, Philippa},
	month = jun,
	year = {2025},
	pages = {186:970--186:992},
}

@article{jung_miri_2026,
	title = {Miri: {Practical} {Undefined} {Behavior} {Detection} for {Rust}},
	volume = {10},
	shorttitle = {Miri},
	url = {https://dl.acm.org/doi/10.1145/3776690},
	doi = {10.1145/3776690},
	number = {POPL},
	urldate = {2026-01-17},
	journal = {Artifact for Miri: Practical Undefined Behavior Detection for Rust},
	author = {Jung, Ralf and Kimock, Benjamin and Poveda, Christian and Muñoz, Eduardo Sánchez and Scherer, Oli and Wang, Qian},
	month = jan,
	year = {2026},
	pages = {48:1383--48:1411},
}

@inproceedings{vanhattum_verifying_2022,
	address = {New York, NY, USA},
	series = {{ICSE}-{SEIP} '22},
	title = {Verifying dynamic trait objects in rust},
	isbn = {978-1-4503-9226-6},
	url = {https://dl.acm.org/doi/10.1145/3510457.3513031},
	doi = {10.1145/3510457.3513031},
	urldate = {2026-01-17},
	booktitle = {Proceedings of the 44th {International} {Conference} on {Software} {Engineering}: {Software} {Engineering} in {Practice}},
	publisher = {Association for Computing Machinery},
	author = {VanHattum, Alexa and Schwartz-Narbonne, Daniel and Chong, Nathan and Sampson, Adrian},
	month = oct,
	year = {2022},
	pages = {321--330},
}

@article{tas_towards_2026,
	title = {Towards verifying unsafe {Rust} programs against {Rust}'s pointer-aliasing restrictions: {Technical} {Report}},
	copyright = {Creative Commons Attribution 4.0 International},
	shorttitle = {Towards verifying unsafe {Rust} programs against {Rust}'s pointer-aliasing restrictions},
	url = {https://zenodo.org/doi/10.5281/zenodo.19254887},
	doi = {10.5281/ZENODO.19254887},
	language = {en},
	urldate = {2026-03-27},
	publisher = {Zenodo},
	author = {Tas, Wannes and Jacobs, Bart},
	month = mar,
	year = {2026},
}

\end{document}